\documentclass[12pt,preprint]{aastex}
\usepackage{epstopdf}

\shorttitle{Galactic Cepheids' infrared emission}
\shortauthors{Barmby et al.}

\begin{document}

\title{Galactic Cepheids with {\em Spitzer}: II. Search for Extended Infrared Emission}

\author{
P. Barmby\altaffilmark{1,2}, 
M. Marengo\altaffilmark{3,2}, 
N.R. Evans\altaffilmark{2},
G. Bono\altaffilmark{4,5},
D. Huelsman\altaffilmark{2,6},
K.Y.L. Su\altaffilmark{7},
D.L. Welch\altaffilmark{8}, 
G.G. Fazio\altaffilmark{2}
}
\altaffiltext{1}{Dept. of Physics and Astronomy, University of Western Ontario, London, Ontario, N6A 3K7 Canada}
\altaffiltext{2}{Harvard-Smithsonian Center for Astrophysics, Cambridge, MA 02138}
\altaffiltext{3}{Dept. of Physics and Astronomy, Iowa State University, Ames, IA 50011}
\altaffiltext{4}{Dept. of Physics, Universit\`a di Roma Tor Vergata,  via della Ricerca Scientifica 1, 00133 Roma, Italy} 
\altaffiltext{5}{INAF--Osservatorio Astronomico di Roma, via Frascati 33,  00040 Monte Porzio Catone, Italy}
\altaffiltext{6}{University of Cincinnati, Cincinnati, OH, 45219}
\altaffiltext{7}{Steward Observatory, University of Arizona, 933 N. Cherry Avenue, Tucson, AZ 85721}
\altaffiltext{8}{Dept. of Physics and Astronomy, McMaster University,  Hamilton, Ontario, L8S 4M1, Canada} 

\begin{abstract}
A deep and detailed examination of 29 classical Cepheids with the {\em Spitzer\/}{} Space Telescope 
has revealed three stars with strong nearby extended emission detected in multiple bands which appears
to be physically associated with the stars.  RS~Pup was already known 
to possess extended infrared emission, while the extended emission around the other two stars 
(S~Mus and $\delta$~Cep) 
is newly discovered in our observations. Four  other stars (GH~Lup, $\ell$~Car, T~Mon and X~Cyg) show 
tentative evidence for extended infrared emission. An unusual elongated extended object next to SZ~Tau 
appears to be a background or foreground object in a chance alignment with the Cepheid. 
The inferred mass loss rates upper limits for S~Mus and $\delta$~Cep are in the range from 
$10^{-9}$ to $10^{-8}$~M$_\odot$~yr$^{-1}$, with the upper limit for RS Pup as high as 
$10^{-6}$~M$_\odot$~yr$^{-1}$. Mass loss during 
post-main-sequence evolution has been proposed as a resolution to the discrepancy between 
pulsational and dynamical masses of Cepheid variable stars: dust in the lost material would 
make itself known by the presence of an infrared bright nebula, or unresolved infrared excess. 
The observed frequency of infrared circumstellar emission ($< 24$\%) and the mass loss rate we estimate 
for our sources shows that dusty mass loss can only account for part of the Cepheid mass loss discrepancy. 
Nevertheless, our direct evidence that mass loss is active during the Cepheid phase is an important confirmation
 that these processes need to be included in evolutionary and pulsation models of these stars, and 
 should be taken into account in the calibration of the Cepheid distance scale.
\end{abstract}


\section{Introduction}
\label{sec:intro}

Classical Cepheid variable stars are of key importance in the extragalactic distance
scale and in studies of Galactic stellar and chemical evolution.
But the very property that makes them useful as distance indicators---pulsation---also 
makes Cepheids difficult to model accurately. 
For several decades, a major problem in the understanding of Cepheids has been the
discrepancy between masses estimated from stellar evolution theory and
those estimated from pulsation theory \citep{christy68,stobie69,fricke72}. 
While the problem was
partly addressed by revision of radiative opacities and changes in evolutionary tracks
\citep{moskalik92},  the 
discrepancy persists at the 10--20\% level \citep{caputo05, keller06}.
Measurements of several Cepheid masses are now available from binary
systems \citep{benedict07,evans08,evans09},  but the comparison
with evolutionary and pulsational masses has not yet produced a definitive result.

There are a number of possible solutions to the `Cepheid mass discrepancy,' summarized
by \citet{bono06}. Recent studies have come to differing conclusions on which might
be responsible: \citet{caputo05} suggested that mass loss can account for the
discrepancy, while \citet{keller08} concluded that additional mixing in Cepheid main sequence
progenitors was responsible. Mass loss remains one of the less-understood
parameters in stellar evolutionary theory \citep{willson08,vink08}, and its relevance to Cepheid masses
is still not completely clear. In the last several years, several lines of investigation
have suggested that mass loss may be important in Cepheids. These include
the discovery of extended near-infrared emission from a number of Galactic Cepheids 
\citep{merand06,merand07,kervella06}, possibly from circumstellar shells, as well
as theoretical modeling of pulsation-driven mass loss \citep{neilson08}
and its application to Magellanic Cloud Cepheids \citep{neilson09a,neilson10}.
Both the presence of shells around nearby Cepheids  and the evidence for infrared
excess in LMC Cepheids have implications for the stars'
derived diameters and therefore for the distance scale. Systematic effects on distances
are possible, if Cepheid mass-loss is substantial enough
to affect spectral energy distributions, and if it works differently in nearby and distant galaxies 
(e.g., is a function of metallicity).
\citet{neilson10} concluded that mass loss in Cepheids ``has an important effect on the 
structure of the IR P-L [infrared period-luminosity] relations'' although the extent of 
the effect is not well characterized.

Attempts to estimate Cepheid mass-loss rates through infrared observations
began  about 25 years ago. \citet{mcalary86} used
infrared excesses inferred from the IRAS Point Source Catalog to
estimate mass loss rates of $\dot{M} <10^{-9}-10^{-8}$~M$_\sun$~yr$^{-1}$,
while \citet{deasy88} combined
ultraviolet and infrared observations to estimate upper limits of
$\dot{M}<10^{-10}-10^{-7}$~M$_\sun$~yr$^{-1}$. \citet{welch88} used radio
observations to place similar upper limits on the mass loss  in ionized gas.
In general these rates are too low to account for the mass discrepancy. 
Recently, the {\em Spitzer Space Telescope\/} has been used in a number of studies
of Cepheids in the mid-infrared, including
the  period-luminosity relation  \citep{ngeow08,freedman08,marengo10a}.
\citet{kervella09} combined ground-based infrared imaging and interferometric measurements 
with {\em Spitzer\/} data (including data from the program described here)
to detect compact circumstellar envelopes around both $\ell$~Car and RS~Pup; the
latter was previously known to be associated with a reflection nebula.

To investigate the possibility of infrared excesses due to mass loss, 
we have obtained {\em Spitzer\/}
\citep{werner04} observations of a sample of 29 Classical Cepheids.
 Our results on the mid-infrared period-luminosity 
and period-color relations for Cepheids are presented in \citet[Paper~{\rm I}]{marengo10a};
a detailed analysis of extended emission near \object[delta Cep]{$\delta$~Cep} is presented
by \citet{marengo10b}. In Paper~I we determined that there was no
evidence from IRAC colors for warm circumstellar dust. The present
paper investigates the presence of cooler dust around this sample
of Cepheids via a search for extended mid-infrared emission. 
An important consideration is that extended emission may 
be the result of recent star-forming activity near a Cepheid rather than mass loss from the Cepheid itself.
However, distinguishing those two different causes  is not straightforward, so a search
for extended emission places an upper limit on the amount of mass loss.
Except where mentioned otherwise, in all of the following analysis we use the stellar distances given by \citet{fouque07}.

\section{Observational data}
\label{sec:obs}

The sample selection and first-epoch observations are described in detail by \citet{marengo10a}.
Briefly, 29 Galactic Cepheid variables and 3 non-variable supergiants
were observed with both IRAC \citep{fazio04} and MIPS \citep{rieke04}
as part of {\em Spitzer\/}{} Space Telescope General Observer, program ID 30666. 
The IRAC observations described by \citet{marengo10a}
were primarily made in subarray mode. The 6 brightest stars were observed in
full-array mode so that fluxes could be measured by fitting the
outer part of the point spread function (PSF); 4 stars were observed in both full-array and subarray modes.
The MIPS observations used the Photometry Astronomical Observing Template
with the shortest possible observation time (48.2~s) at 24~\micron{}
and varying times  (37.7--1300.2~s) at 70~\micron. Three stars
(S~Nor, V~Cen, U~Sgr) were observed in two adjacent fields of view 
at 70~\micron{} to better cover the surrounding clusters.
While our target SZ~Tau (see \S\ref{sec:sztau}) may also be in the cluster NGC~1647 \citep{turner92}, after inspecting
its 24~\micron{} image, we changed the planned 70~\micron{} observation 
to use only one field of view centered on the star. This allowed us to
increase the observation time and, as a bonus, also provided a second
epoch of 24~\micron{} observations.
We performed simple aperture photometry on post-BCD mosaics primarily from
the pipeline version S16.1 output. For the three cluster stars, the 70~\micron{} data
were remosaiced using MOPEX to combine the data for the two fields of view.
For two stars (RS~Pup, GH Lup) the 70~\micron{} mosaics were remade
to remove negative sidelobes.

\subsection{New IRAC data and reduction}
\label{sec:irac_obs}

To follow-up earlier results, we obtained some additional {\em Spitzer\/} data
in Cycle 5, on the dates 2008 October 5--6. All stars were observed in full-array mode with IRAC
as part of Guaranteed Time program 50350 (PI G. Fazio).  
This increased the field of view for stars which had
previously  been observed only in subarray mode, and also
provided deeper observations (2-second frames at each of 36 dither
positions) for all stars.  
Each target fainter than $K=2$ was observed with the IRAC full frame 12 second High Dynamic Range (HDR) mode, 
using a 12 point Reuleaux small scale dither pattern for a total integration time of 144 sec in each band. 
HDR 12 second frames were used to better 
reveal the area closer to the saturated cores of the targets, which also helped the PSF subtraction process. 
For the six  brightest sources, in order to avoid excess saturation, targets were observed with 2 second frames using IRAC full frame 
non-HDR mode and the 36 point Reuleaux small scale dither pattern, for a total integration time of 72~s in each band. 
The Reuleaux dither pattern  was used to achieve the required total exposure time while obtaining the best possible 
spatial sampling of the targets in anticipation of the PSF subtraction process. 

The basic calibrated data (BCD) for all observations were reduced using the custom post-BCD software IRACproc 
\citep{schuster06}, which generated mosaics for each source with the final exposure depth and outliers removed. 
The final images had a pixel scale of 0.8627 arcsec~pix$^{-1}$ ($1/\sqrt{2}$ the original IRAC pixel scale) which 
provides a better sampling of the sources' PSF  for a more accurate PSF subtraction. The PSF subtraction was 
also performed using a routine part of IRACproc, which has been extensively tested in other \textit{Spitzer} 
programs \citep{marengo06,marengo10a,luhman07} to search for faint structures and point sources around bright stars.

\subsection{MIPS photometry}
\label{sec:mipsphot}

Although the MIPS observations used in this paper are
the same as those in \citet{marengo10a}, a major purpose
of the present work is to examine extended emission around 
a subset of the stars in our sample. Accordingly,
we have performed point source fitting and subtraction on
the 24~\micron{} images of some targets 
to separate the contributions of the star and extended
emission to the total flux density. 
The data were reprocessed with the MIPS instrument team Data Analysis Tool using
standard procedures. The PSF used for the 24 $\mu$m subtraction was built from the
observations of the A4 V star, $\tau_3$ Eri \citep{su08}, which matches
well with the theoretical PSF computed from STinyTim Program
\citep{krist06} after proper smoothing. The subtraction was done by
iteratively shifting the PSF to match the centroid of the target in sub-pixel accuracy, 
adjusting its scaling, and subtracting it from the target image until the residual 
inside $r<12$\arcsec~($\sim$ to the first bright Airy ring of the PSF) was minimized. 
The residual flux is generally less than 1\% of the total flux of a typical point source on a clean background.

To estimate the non-point source emission near stars showing extended
emission in the MIPS bands, 
we used the results of point source fitting and subtraction. We performed aperture photometry
of the point and extended emission on images which had the field  (but
not the target) stars subtracted, using the largest apertures which
could be contained in the image. From this we subtracted the flux measured by
fitting and subtracting the target stars to derive the total non-point source emission.
This procedure depends on accurate background subtraction for the large-aperture
photometry, particularly difficult in the 70~\micron{} images, and the resulting
uncertainties are large. 
An alternative approach which can constrain the physical
conditions within regions of extended emission is used by \citet{marengo10b}:
measuring the surface brightness of the extended emission in small areas. 
Because the stars are heavily saturated in the IRAC bands, this
is the only method available to investigate the extended emission visible
in the IRAC images. In both IRAC and MIPS images we measured the surface brightness in several 
small regions near stars with extended emission, and subtracted backgrounds
measured in regions located off the extended emission. The regions used are shown in the images of
the relevant stars; the measured surface brightnesses
are reported in Table~\ref{tab:surfphot}.

\section{Analysis}
\label{sec:analysis}

Infrared colors were used in Paper~I to demonstrate a lack of warm 
circumstellar dust in our sample of Cepheids.
However, dust mass lost from Cepheids and then transported to larger distances would
be expected to be cooler, and possibly visible as extended infrared emission.
One example of this phenomenon is the infrared nebula visible around
$\delta$ Cep \citep{marengo10b}: here we complete the survey of our Cepheid and comparison star
sample. None of the stars showed any extended emission in the 3.6 $\mu$m or 4.5 $\mu$m bands, 
so those bands are not discussed further here.
Distinguishing between extended emission associated with a specific star
and that due to nearby Galactic cirrus and/or star formation, especially at low Galactic latitudes, is not always straightforward.
We expect that emission associated with an individual star should be centered on  or near the star, 
with its surface brightness decreasing with increasing distance from the star. 
Unrelated  infrared emission will not in general have these characteristics, although of course 
it could hide or confuse real low level emission. For brevity,
we refer to unrelated infrared emission as `cirrus' below, although we recognize that it may not be
true Galactic cirrus emission.

\begin{figure}
\includegraphics[width=18cm]{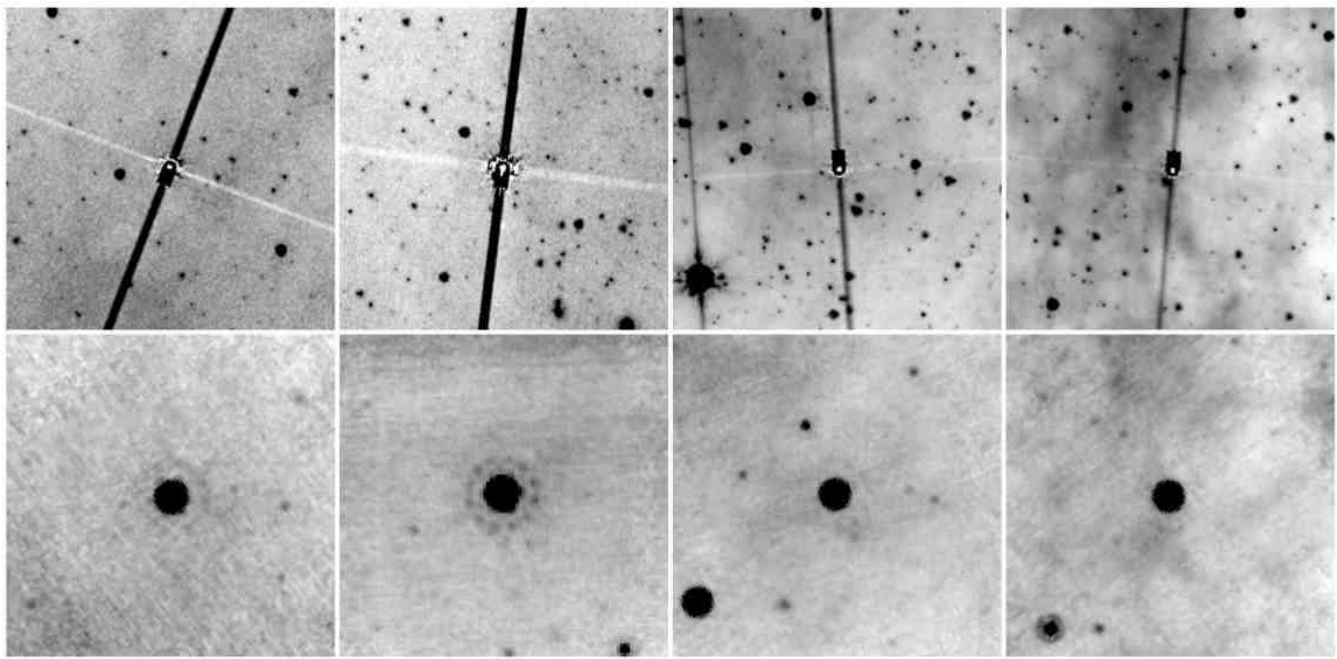}
\caption{
Images of some Cepheids with nearby  extended
infrared emission unrelated to the star.
Left to right:  DT~Cyg, FF Aql, V~Cen, and VY~Car.
Top row: PSF-subtracted IRAC 8.0~\micron{} images,
Bottom row: non-subtracted MIPS 24~\micron{} images.
All fields of view are 4 arcminute squares.
In these and all following images, north is oriented up and east left.
The black and white bands in the IRAC images are artifacts from both the bright source
and the PSF-subtraction process.
}
\label{fig-cirrus}
\end{figure}

Of our sample of 29 Cepheids, 16 showed clear signs of  cirrus in the
IRAC 8~$\mu$m images: 
\object{AQ Pup}, \object{BF  Oph}, \object{DT Cyg}, \object{FF Aql}, \object{GH Lup}, \object{RS Pup}, \object{S Mus}, \object{S Nor},  
\object{SW Vel}, \object{T Mon}, \object{U Car}, \object{U Sgr}, \object{V636 Sco}, \object{V Cen},  \object{VY Car}, and \object{X Cyg}. 
Cirrus is also visible in the MIPS images of GH~Lup, V Cen, VY~Car and U~Sgr.\footnote{Of our 
three non-variable comparison stars, \object{HD~1822296} and \object{HD~183864} were observed in subarray mode only; 
\object[psi And]{$\psi$ And} was observed in full-array mode and shows no evidence for cirrus.}
As expected, the presence of cirrus is related to Galactic latitude: 
of the Cepheids in our sample with $|b|<10^{\circ}$, 15 of 23 appear to
be surrounded by cirrus emission. The remaining star with apparent unrelated
emission is DT~Cyg, at $b=-10.8^{\circ}$.
The list of cirrus sources include three of the four stars we believed 
as cluster members (S~Nor, V~Cen, U~Sgr); this classification is confirmed by \citet{gieren97}, who
also found that VY~Car is a member of the Car~OB~2 association.
The extended emission near these stars could be associated with either their
clusters or the general interstellar medium.
Figure~\ref{fig-cirrus} shows some examples of stars surrounded by  cirrus.

As a quantitative test for excess infrared emission near the stars, we computed the average
surface brightness around each star and compared it to a background measurement. The IRAC 
8~\micron{} images were used for this measurement as they have a larger uniform-exposure field
of view than the MIPS images, and much better sensitivity to extended emission than the 5.8~\micron{} images.
The average surface brightnesses were measured in  1 arcminute radius apertures surrounding each star,
with the PSF subtraction residuals and bright star artifacts masked. The background measurement
was made in an annulus with inner and outer radii of 74 and 98 arcseconds.
Point sources near the target stars, imperfect PSF subtraction, and unrelated cirrus emission
all complicate the comparison between low-surface-brightness emission near the stars
and in their vicinities. To be conservative, we consider as having possible extended infrared
emission only those stars for which the average surface brightness in the measurement
aperture exceeds that in the background aperture by 20 times the quadrature sum of
the surface brightness  standard deviations. Since our measurement
errors are  are influenced by the factors listed above and are unlikely to be Gaussian, 
we do not consider these detections to have true $20\sigma$ significance---the value chosen is
intended as a starting point to guide the analysis.

We found no evidence for extended infrared emission 
around two stars which have been previously found to have circumstellar envelopes:
\object{Polaris} \citep{merand06} and \object{Y~Oph} \citep{merand07}.\footnote{The
extended emission found by those authors was at  very small spatial scales (a few stellar radii), well below our angular resolution.}
Other stars with neither cirrus or extended infrared
emission are \object{BB Sgr}, \object{U Aql}, \object{V350 Sgr}, \object{W Sgr}, \object[beta Dor]{$\beta$ Dor}, \object[eta Aql]{$\eta$ Aql}
and \object[zeta Gem]{$\zeta$ Gem}.
Five stars showed both cirrus and extended emission centered on or near the star:
GH Lup, RS Pup, S Mus, T Mon, and X Cyg. Three further stars had little detectable cirrus but did have
extended infrared emission centered on the star: $\delta$ Cep, $\ell$~Car, and SZ Tau.

The extended emission around $\delta$ Cep is discussed in detail by \citet{marengo10b} and briefly summarized here.
The emission consists of a large-scale (0.1~pc) arched structure visible at 70~\micron{},
with diffuse, filamentary emission at 5.8, 8.0 and 24~\micron{} contained within the arch.
The brightest region of this diffuse emission is between $\delta$~Cep and its hot companion \object{HD 213307}.
The symmetry axis of the 70~\micron{} emission near $\delta$~Cep is aligned with the
star's velocity vector relative to the local ISM, suggesting a bow-shock at least partly
swept-up from the ISM. The emission near the stars is hypothesized to be due
to the interaction of their stellar winds. The mass loss rates required for the creation of these 
structures are estimated to be in the range $5\times10^{-9} - 6 \times 10^{-8}$~M$_{\sun}$~yr$^{-1}$.

The remaining stars with extended infrared emission are discussed individually below. 

\subsection{GH Lup}

\begin{figure}
\includegraphics[width=18cm]{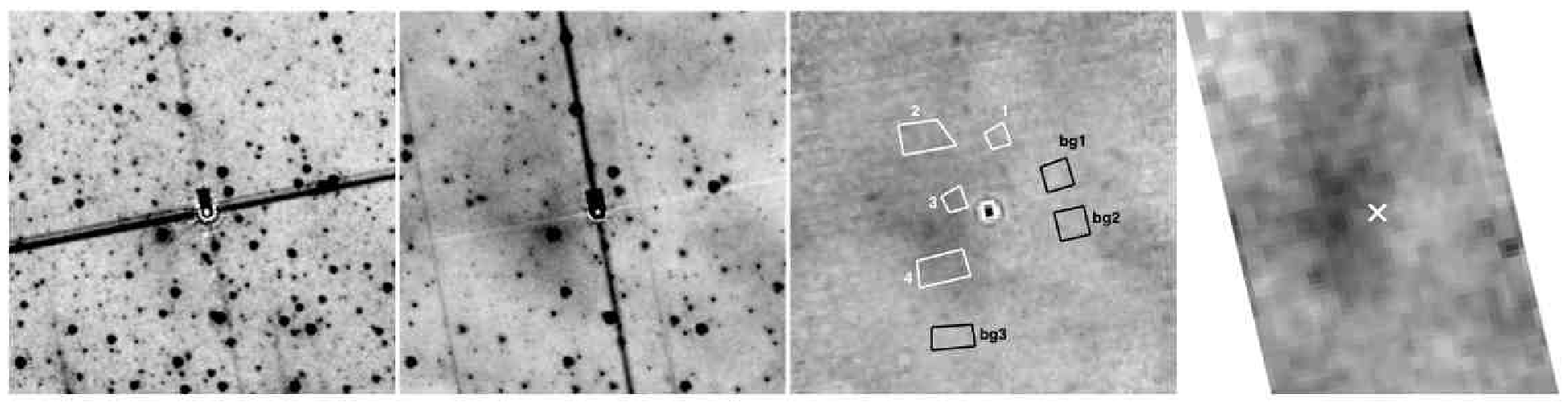}
\caption{
PSF-subtracted images of GH Lup in the 
(left to right) 5.8 $\mu$m and 8.0 $\mu$m IRAC bands,
and the 24 $\mu$m  and  70 $\mu$m (not PSF-subtracted; the white X marks the position
of the star) MIPS bands. The field of view is a 4 arcminute square. 
Numbered boxes are regions used for surface photometry (see \S\ref{sec:mipsphot}).
}
\label{fig-GHLup}
\end{figure}

\begin{figure}
\includegraphics[width=12cm]{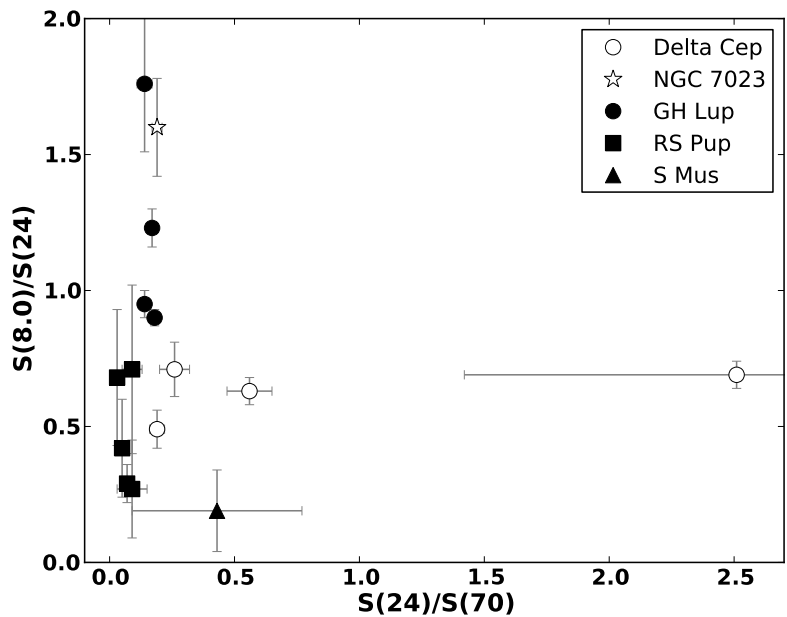}
\caption{
Left: Surface brightness ratios of extended emission near GH Lup, RS~Pup and
S~Mus, with emission near $\delta$~Cep (circles) and NGC~7023 (star) shown for comparison. 
Increasing values on the horizontal axis correspond roughly to increasing dust temperature,
while increasing values on the vertical axis correspond to increasing PAH content.
}
\label{fig:sb_ratios}
\end{figure}

Images of \object{GH Lup} are shown in Figure \ref{fig-GHLup}. The most obvious component
of extended emission is an arc-like structure to the east of the star, visible at
8, 24, and 70~\micron{}. The structure is not centered on the star itself but roughly 
20\arcsec\ to the south ($2.2\times10^4$~AU at the star's distance) with a radius
of approximately 32\arcsec\ ($3.6\times10^4$~AU).  Spatially coincident with the
arc is a point source visible at 5.8 and 8.0~\micron{} only, which we can identify with
the $K_s = 9.5$ object \object{2MASS J15244147-5251282}. Because the region is crowded,
with copious diffuse emission, it is unclear whether the arc near GH~Lup is related to the
Cepheid, the nearby star, or neither. The stellar crowding also makes defining regions for
surface photometry of the diffuse emission difficult; we chose the regions marked
in Figure~\ref{fig-GHLup} to avoid stars in the 8~\micron{} image and capture
at least some of the extended emission. The resulting measurements are given in
Table~\ref{tab:surfphot}, with the colors of the extended emission plotted in 
Figure~\ref{fig:sb_ratios}.
Using a modified blackbody spectrum to assign color temperatures from
the $S_{24}/S_{70}$ ratios yields $55<T<81$~K, depending on the emissivity index $\beta$ for the modified blackbody. 
The $S_{8}/S_{24}$ ratios for GH Lup tend to be higher than those found for
the other stars with extended emission. Together with the structure of the emission and
the extensive nearby cirrus, this leads us to speculate that the emission 
near GH Lup is only moderately likely to be associated with the star itself (see \S\ref{sec:disc}).

\subsection{$\ell$ Car}

\begin{figure}
\includegraphics[width=18cm]{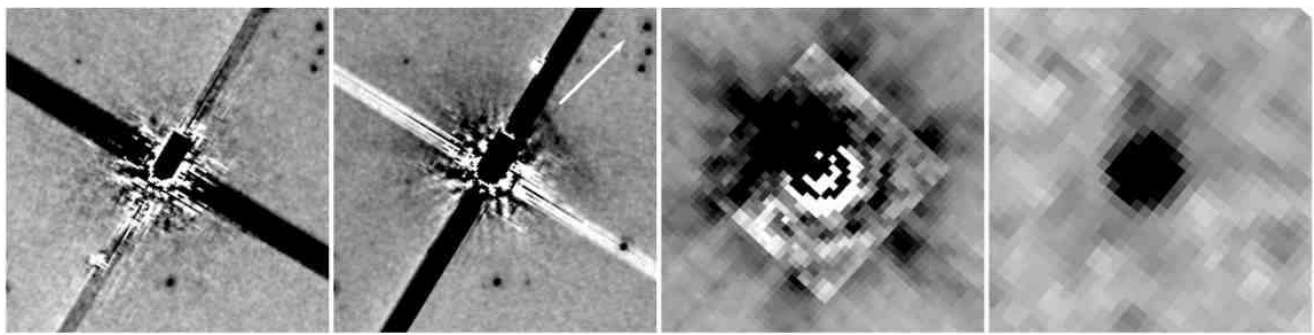}
\caption{
PSF-subtracted images of $\ell$~Car in the 
(left to right) 5.8 $\mu$m and
8.0 $\mu$m IRAC bands,
and the 24 $\mu$m  and 
70 $\mu$m (not PSF-subtracted) MIPS bands. The field of view is a 2 arcminute square. 
The white arrow in the second panel shows the direction of the star's proper motion relative to the local ISM.
}
\label{fig-l_Car}
\end{figure}

Images of \object[l Car]{$\ell$ Car} are shown in Figure \ref{fig-l_Car}. 
These data were also analyzed by \citet{kervella09}, who failed to detect any nebulosity in the {\em Spitzer}{} images, but found 
evidence of small scale ($\sim1$\arcsec) extended emission in near- and mid-IR interferometric data, attributed to hot circumstellar dust. 
Our PSF-subtracted IRAC images show low surface brightness emission, at 5.8 and 8.0~\micron, surrounding the star with a 
radius of $\sim 35$\arcsec, or $1.7\times10^4$~AU at a distance of 498~pc \citep{kervella09}. 
We do not detect any significant 
extended emission at 70~\micron, and only a tentative excess at 24~\micron, northeast of the nearly saturated core of the PSF-subtracted star. 
The 8.0~\micron{} emission appears brighter northwest of the star, 
where an arc-like structure is detected at a distance of  $\sim 30$\arcsec{} ($1.5\times10^4$~AU).  
Following a suggestion by the referee, we computed the components of the star's
Galactic peculiar space motion and their projection onto the sky, using the procedure described by \citet{marengo10b}. The
star's position and proper motion were taken from the Hipparcos catalog \citep{perryman97},
the distance from \citet{kervella09}, and heliocentric radial velocity of 3.6~km~s$^{-1}$ from \citet{nardetto09}. 
We find a space velocity of 24.9~km~s$^{-1}$ along a position angle of 315\arcdeg\ (east of north), with
the direction shown as an arrow in the second panel of  Figure \ref{fig-l_Car}. 
The vector is roughly aligned along the direction from the star to the arc-like structure, 
suggesting a possible connection between the two (e.g., a bow shock, reflection nebula, or light echo).
Image artifacts from the PSF subtraction and the star make the connection difficult to secure, however.
Artifacts also make it impossible to accurately measure the surface brightness of the extended emission, 
and thus estimate the mass loss rates that could be responsible for the observed nebulosity.
Higher spatial-resolution imaging covering a wider area would clearly be useful in further study of $\ell$ Car.

\subsection{RS Pup}

\begin{figure}
\includegraphics[width=18cm]{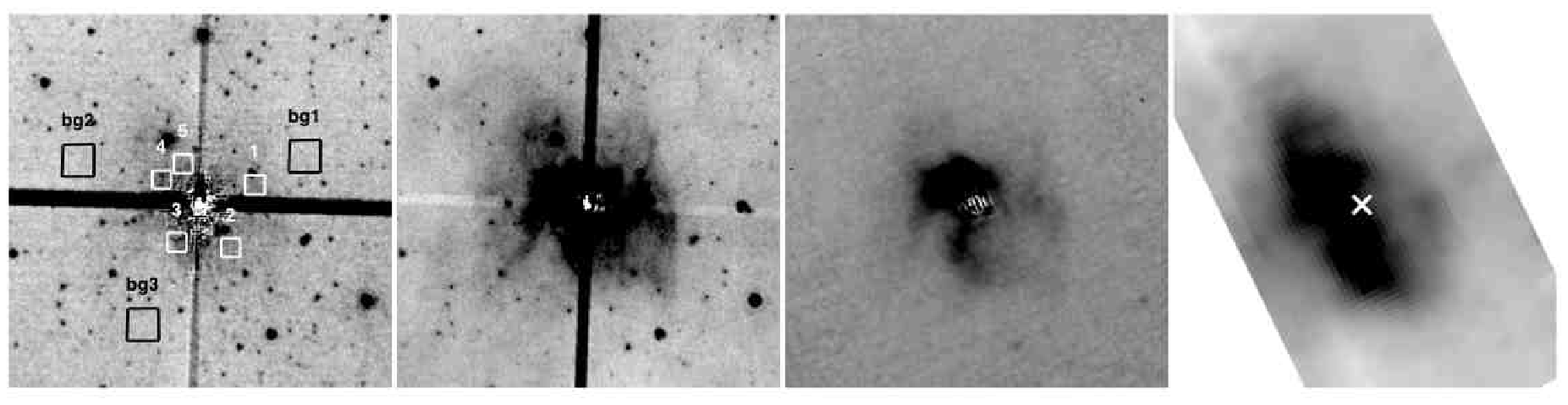}
\caption{
PSF-subtracted images of the reflection nebula around RS Pup in the 
(left to right) 5.8 $\mu$m and 8.0 $\mu$m IRAC bands,
and the 24 $\mu$m  and 70 $\mu$m  MIPS bands (not PSF-subtracted; the white X marks the position
of the star).
The field of view is a 4 arcminute square.
Numbered boxes are regions used for surface photometry (see \S\ref{sec:mipsphot}).
}
\label{fig-RS_Pup}
\end{figure}

\object{RS Pup} is embedded in a well-known reflection nebula \citep{westerlund61,havlen72} whose infrared emission has been
noted by numerous authors \citep[e.g.,][]{gehrz72,mcalary86}. 
PSF-subtracted images of RS~Pup are shown in Figure \ref{fig-RS_Pup}. The
morphology of the extended infrared emission varies with wavelength but
the brightest region is to the northeast of the star. Most of the emission is within
about 80\arcsec\ of the star, a linear distance of 0.77~pc at the 1.992~kpc distance of RS~Pup \citep{kervella08}, 
although the small field of view of the 70~\micron\ image makes this difficult to pin down.
The MIPS data, but not the full-frame IRAC data, were also analyzed by \citet{kervella09}, so
here we present the first image of the resolved 8.0~\micron\ emission around RS~Pup.
\citet{kervella09} attributed the $\gtrsim 20$~\micron\ resolved emission, as well as the
infrared excess in the IRAS data, to an extended envelope with a radius of roughly 1~pc
and a temperature of $\sim40$~K.
\citet{kervella09} also detected resolved emission at much smaller angular distances 
to the star, and attributed this to a ``warm and compact component'' at a distance of 300--3000~AU
and a temperature of a few hundred K. They proposed that the warm component comes
from RS Pup's mass loss, and the cold component is interstellar medium gas compressed by 
the star's stellar wind. In \citet{marengo10b} a similar origin for the extended infrared
emission around $\delta$~Cep was proposed.

We add to the analysis of RS~Pup's extended emission by examining the colors of the
extended emission. Using the surface brightness measurements in Table~\ref{tab:surfphot}, 
Figure~\ref{fig:sb_ratios} plots the ratios of the extended emission
around RS~Pup\  and S~Mus, with the values derived for $\delta$~Cep by \citet{marengo10b} for comparison. 
Using a modified blackbody spectrum to assign color temperatures from
the $S_{24}/S_{70}$ ratios yields $50<T<80$~K, depending on the emissivity index $\beta$ for the modified blackbody. 
These temperatures are a little lower than the values derived for $\delta$~Cep; however, compared to $\delta$~Cep, the RS~Pup nebulosity is more
distant from the central star. 
Like the $\delta$~Cep nebulosity, the RS Pup extended emission has $S_{8.0}/S_{24}\sim 0.5$, suggesting 
that some polycyclic aromatic hydrocarbons (PAHs) must be present 
\citep[pure dust at $50<T<80$~K would imply a much lower ratio;][]{marengo10b}.
We can also compute an average dust temperature for the RS Pup nebulosity
using the total flux densities of the extended emission. At 70~\micron{} we
used the largest aperture that will fit on the image (85\arcsec\ radius) to derive
a total flux density of 9.7~Jy. At 24~\micron{}  we measured the flux density in
a large aperture (72\arcsec{} radius) and subtracted the PSF-fitting flux of
the central source to derive an extended emission flux density of ($620-330=$)~290~mJy.
Using a modified blackbody spectrum to assign color temperatures from
the ratio of the two bands yields $44<T<57$~K, depending on $\beta$. 
These temperatures are a little higher than the estimate of 40~K given by
\citet{kervella09}, who extrapolated the 70~\micron{} surface brightness profile
to derive a larger total flux density.

The dust mass in the RS~Pup nebula is best estimated with far-infrared photometry,
since as Figure~\ref{fig-RS_Pup} shows, the stellar emission is insignificant
at wavelengths of 70~\micron{}  and beyond. Because our 70~\micron{} MIPS images 
have a rather small field of view, the difficulty of accurate background subtraction makes 
any attempt to estimate the total flux density \citep[including that of][]{kervella09} problematic. 
However, a rough estimate  of total dust mass is possible and provides
a comparison point for previous results.
We use the formula given in Eq.\ 2 of  \citet{evans03},  assuming that all of the 70~\micron{} emission
is from optically thin dust with $T=40-80$~K and emissivity $\kappa_{\nu}=3Q_{\nu}/\rho a$ in cm$^2$~g$^{-1}$.
For comparison with \citet{mcalary86}, we use $\kappa=19.0$, scaled from the 100~\micron{} value 
the $a=0.1$~\micron{} `dirty silicate' dust model in \citet{jones76} via a
$Q_{\nu} \propto \lambda ^{-2}$ wavelength dependence.
The derived total dust masses for RS Pup are in the
range $1\times10^{-3}-1.4\times10^{-2}$~M$_{\sun}$. 
(The range of values here accounts for the change in the Planck function, but
not any temperature dependence of the dust emissivity, over the given temperature range.)
The uncertainty due to temperature is compounded by the uncertainty in
dust emissivity: for example, using the wavelength-dependent emissivity of
amorphous fayalite given by \citet{evans03} would result in a value of $\kappa$ nearly
15 times larger (and thus masses 15 times smaller) than given above.
The 100~\micron{} flux density from the IRAS Small Scale Structure Catalog \citep{helou88}
should provide a better estimate of the total flux density of the RS~Pup nebula.
Using this measurement, \citet{mcalary86} estimated a total dust mass of 0.028~M$_{\sun}$ (for their adopted distance of 1.778 kpc), 
or 0.035~M$_{\sun}$ for the 1.992 kpc distance we use. The mass derived from the 70~\micron{} flux density is 
the same order of magnitude, likely smaller because our
observations do not capture the full extent of the RS Pup nebula.

As \citet{mcalary86} point out, the large size of
the RS Pup nebula makes estimating a mass-loss rate from the nebular mass ``extremely difficult''.
With a gas-to-dust ratio of 100, the \citet{mcalary86} dust mass converts to a total mass of 3~M$_{\sun}$.
An argument in favor of the idea that some of the nebula is swept-up interstellar medium (ISM)
results from a simple comparison of the nebula mass with the helium-burning timescale
of RS~Pup \citep{kervella09}. Taking the mass estimate for RS Pup provided by \citet{caputo05}, 
the helium burning time inside the instability strip is of the order of 1.2 Myr.
If all of the nebula consists of mass lost by RS~Pup during its helium-burning phase,
the implied mass-loss rate would be $\gtrsim10^{-6}$~M$_{\sun}$~yr$^{-1}$. 
This would imply a total mass loss at least of the
order of 10\%. As \citet{kervella09} note, this would be sufficient to solve the
pulsational/evolutionary mass difference for RS Pup. However, as noted by
 \citet{neilson08}, the observational mass loss rate for RS Pup  disagrees with their theoretical
 predictions ($1.6-6.5 \times 10^{-9}$~M$_{\sun}$~yr$^{-1}$) by several orders of magnitude.
These predictions do appear to fit the observational data for Magellanic Cloud Cepheids
\citep{neilson09a,neilson10}, which
makes it more believable that at least some of the mass in the RS Pup nebula did not originally belong to the star.
The  observed $S_{8.0}/S_{24}$ ratio is too high for pure dust emission, but
lower than typical for the typical PAH content of the ISM. Since
Cepheid atmospheres are  oxygen-rich and not expected to form PAHs,
this suggests that both swept-up ISM (or pre-existing material containing PAHs) and stellar wind
must comprise the nebula.

\subsection{S Mus}

\begin{figure}
\includegraphics[width=18cm]{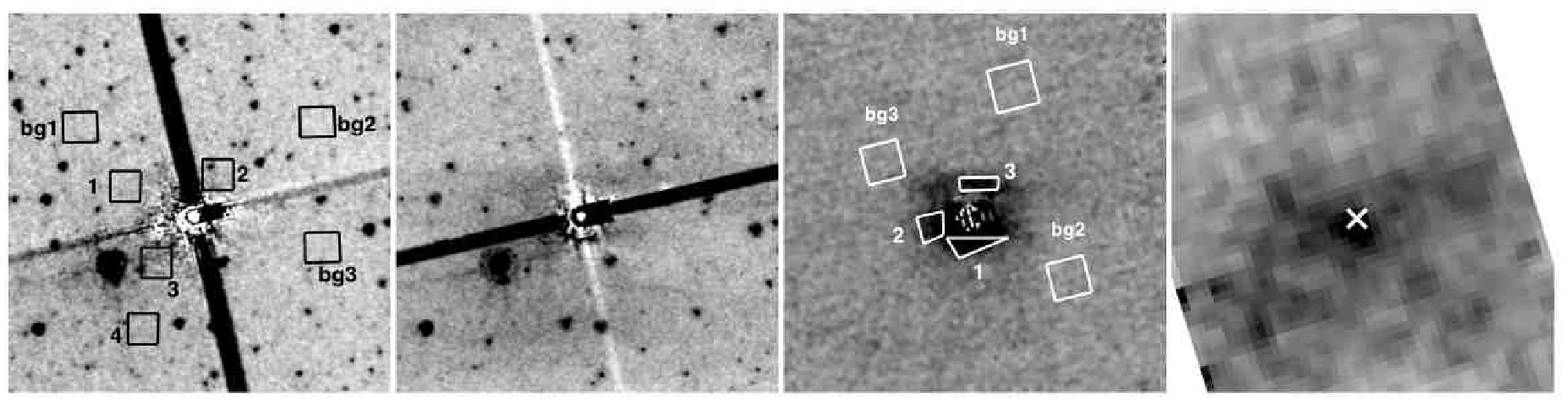}
\caption{
PSF-subtracted images of S Mus in the 
(left to right) 5.8 $\mu$m and
8.0 $\mu$m IRAC bands,
and the 24 $\mu$m  and 
70 $\mu$m  (not PSF-subtracted) MIPS bands; the white X marks the position
of the star.
The field of view is a 3 arcminute square.
Numbered boxes are regions used for surface photometry (see \S\ref{sec:mipsphot}).
}
\label{fig-SMus}
\end{figure}

Images of \object{S Mus} are shown in Figure~\ref{fig-SMus}. There are slightly extended emission features
to both the north and south of the star visible at both 8 and 24~\micron,
with radial extents 35--45\arcsec\ ($2.9-3.7\times10^4$~AU at a distance of 820~pc).
There is also extended emission visible at 70~\micron{}, to a radius of about 55\arcsec{} ($4.5\times10^4$~AU),
although the morphology is different at that wavelength.
The surface brightness of this emission is faint compared to that around RS~Pup, and the
orientation of the PSF subtraction residuals makes computing  $S_{8.0}/S_{24}$ and $S_{24}/S_{70}$
at the same location difficult. The S Mus point shown in Figure~\ref{fig:sb_ratios} is for
the box marked `3' in the IRAC images and `2' in the MIPS images.
Although uncertain, the colors imply an even lower PAH content that in RS~Pup and $\delta$~Cep,
and a somewhat higher color temperature (70--120~K).
This star has a high-temperature companion \citep{evans06} and the extended emission 
could be associated with that star rather than S~Mus itself, or the companion
could be responsible for heating the circumstellar material \citep[as hypothesized for
$\delta$~Cep;][]{marengo10b}.

The photometry of the extended emission around  S~Mus  can be used to estimate
the mass lost by the star. Large-aperture (35\arcsec\ radius) photometry of the 70~\micron{} emission yields
a total flux density of $96\pm 12$~mJy; subtracting the point-source flux density (measured in a 16\arcsec\ radius aperture
as for Paper~I) of $39\pm 11$~mJy yields a nebular flux density of $57\pm 16$~mJy.
We use the same dust mass estimation procedure as for RS~Pup,
a temperatures of 70--120 K, and a stellar distance of 0.88~kpc to derive
an  implied dust mass $4-17\times10^{-7}$~M$_{\sun}$.
With a gas-to-dust ratio of 100, the total implied mass  would be approximately 
$4-17\times10^{-5}$~M$_{\sun}$. 
If we instead use $\kappa=56$ as in \citet{marengo10b}, the total mass agrees
quite well with that derived  $\delta$~Cep.
Given the photometric uncertainties, the agreement may well be coincidental; 
however the order of magnitude comparison is intriguing.
A mass loss timescale can be estimated by dividing the distance to the emission
(35\arcsec\ is 0.15~pc at the distance of S~Mus) by the escape velocity (100 km~s$^{-1}$)
to yield $t\approx1500$~yr, so the resulting mass loss rate would be $3-10\times10^{-8}$~M$_{\sun}$~yr$^{-1}$.
As for $\delta$~Cep, this estimate is consistent with the predictions of \citet{deasy88} but
well above those of \citet{neilson08}. 
There are two reasons why our value for the mass loss rate is likely to be an 
upper limit, and therefore larger than the \citet{neilson08} predictions.
(1) We assumed that all of the dust responsible for the extended
emission comes from wind material as opposed to having been swept-up from the ISM;
as for RS Pup, the PAH content implies that ISM sweeping is likely relevant. 
(2) Using the escape velocity to estimate the nebula's age results in the minimum
possible age, and hence a maximum for the mass loss rate.

We can estimate the total mass lost for S~Mus through its mass and
crossing time for the instability strip, as above for RS Pup. 
We estimated pulsational and evolutionary masses through the relation provided by \citet[][Tables 2--5]{caputo05};
using the near-infrared photometry from \citet[][transformed to the 2MASS system
using \citealt{koen07}]{laney94} and optical photometry from \citet{pedicelli10}, 
corrected for extinction using the reddening given by \citet{fernie95} and the
reddening law of \citet{cardelli89}.
We used an intrinsic distance based on the near-infrared photometry and period-Wesenheit relations
provided by \citet{persson04}. This procedure results in a mean mass $7.0\pm 1.5$~M$_{\sun}$,
and  according to \citet{bono00} the corresponding crossing time of the
instability strip is of the order of 0.14~Myr. Given the mass loss rate above, this
results in a small total mass loss of 0.006~M$_{\sun}$.
Using the recent estimate of the radius of S~Mus by \citet{groenewegen08}
and the period-mass-radius relation given by \citet{bono01a}, the pulsation mass is
lower, $\sim 5$~M$_{\sun}$, and the instability strip crossing time much longer, of order 1~Myr;
however the total mass lost would still be only 0.04~M$_{\sun}$, too small to significantly affect
the mass discrepancy.

\subsection{SZ Tau}
\label{sec:sztau}

\begin{figure}
\includegraphics[width=18cm]{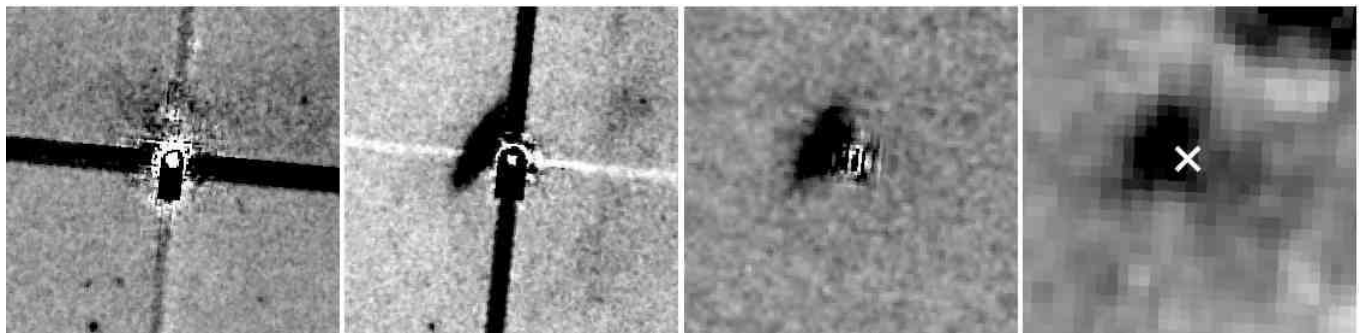}
\caption{
PSF-subtracted images of SZ Tau in the 
(left to right) 5.8 $\mu$m and
8.0 $\mu$m IRAC bands,
and the 24 $\mu$m  and 
70 $\mu$m  (not PSF-subtracted) MIPS bands; the white X marks the position
of the star.
The field of view is a 2 arcminute square.
Of particular interest is the elongated object to the northeast of the star.
}
\label{fig-SZ_Tau}
\end{figure}

\object{SZ Tau} was the only Cepheid found in Paper~I to have a statistically significant $[24]- [70]$ color excess. 
Images of this star are shown in  Figure \ref{fig-SZ_Tau}. 
There is an extended object about $25^{\prime\prime}\times 8^{\prime\prime}$ clearly visible in the 
PSF-subtracted 8.0 $\mu$m image, about   7\arcsec\  to the north-east of the star.
At the 512~pc distance to the star,  $25^{\prime\prime}\times 8^{\prime\prime}$ corresponds
to linear scales of $(1.3 \times0.4)\times 10^4$~AU.
The object  is not clearly visible in the 5.8 $\mu$m band, due to  PSF subtraction residuals, however
it is visible at 24~\micron{} and appears to be the source of most of the 70~\micron{} luminosity.
The object is resolved at 24~\micron{} and has a similar appearance in
two images in that band taken about a year apart, suggesting
that it is not an instrumental artifact or a solar system object.
\citet{noriega-crespo97} detected a ``ridge or filament'' of similar shape but 
much larger angular size in IRAS imaging of Betelgeuse; they suggested that 
this structure is emission from the interstellar medium, unrelated to the star.
\citet{turner92} concluded that SZ~Tau is a first-overtone pulsator and noted
several other unusual characteristics of this star (e.g., period changes),
but it is unclear how these features could be related to the presence of extended emission.

The elongated shape of the emission near SZ Tau suggests a background galaxy. 
The infrared colors should yield further information on the object's nature, although
photometry of the SZ Tau extended emission is complicated by the star itself.
At 70~\micron, the extended emission appears to be the dominant source
of emission. When we measure flux density in a 35\arcsec\ radius aperture
centered on the 70~\micron{} peak and a 39--65\arcsec\ background annulus,
the resulting flux density (which may include some contribution from the star) is 74~mJy.\footnote{The 
flux density for this star reported in \citet{marengo10a} is slightly different because
that value was measured centered on the 24~\micron{} star position.
Also, there is a bright far-infrared source about 80\arcsec\ to the northwest of the star's
position which may also contaminate the photometry.}
 We measured the flux density of the SZ Tau extended emission at 24~\micron{} by 
doing a large-aperture (35\arcsec\ radius) measurement on the extended emission+star
and subtracting the PSF-fitting star flux, to give an extended emission flux density of
($175-144=$) 31~mJy. Measuring the 8~$\micron$ flux density is the most difficult
since there is significant contamination from the PSF subtraction residuals.
Measuring the emission on an image with the residuals masked
in 35\arcsec{} radius aperture with 35--43\arcsec{} background annulus
gives a flux density of 10~mJy; applying the IRAC extended source photometric
correction changes this to 7.4~mJy. Different approaches to residual masking
and photometric aperture change the resulting flux density by $\sim 20-30$\%, so the
value is highly uncertain. This is hardly surprising since the extended emission
is more than 100 times fainter than the star at 8~\micron{}.
Consistent with this result is the classification of the star's mid-infrared
spectrum as normal, with no dust, based on data from the PHT-S instrument on
the Infrared Space Observatory \citep{hodge04}.

\begin{figure}
\includegraphics[width=12cm]{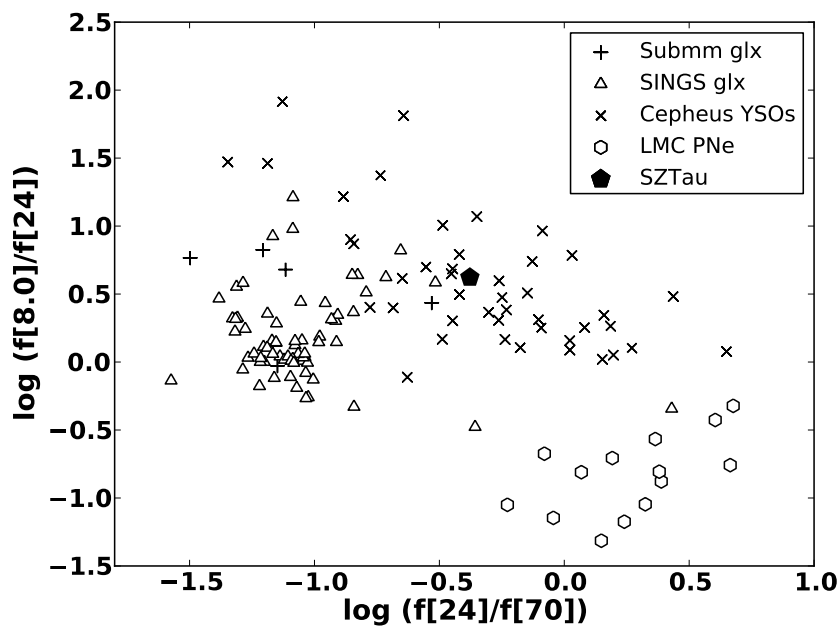}
\caption{
Right: flux density ratios for submillimeter-detected galaxies \citep{hainline09},
nearby galaxies \citep{munoz-mateos09}, Cepheus young stellar objects \citep{kirk09},
LMC planetary nebulae \citep{hora08}, and the extended emission near SZ Tau.}
\label{fig:colors}
\end{figure}

We can ask whether the the galaxy-like appearance of the extended emission near SZ Tau 
is consistent with its mid-infrared flux densities and colors. The object is rather bright
for a galaxy: number counts at 24~\micron{} \citep{shupe08} show that
there are only 3.5--4 galaxies per square degree brighter than about 15~mJy.
Taken together, all of our 24~\micron{} Cepheid images cover about
0.5 square degree, so we might expect 1 or 2 such galaxies in our survey.
Figure~\ref{fig:colors} compares the colors of the SZ Tau extended emission with
some comparison samples. The color $f_{8.0}/f_{24}=0.24$ 
is consistent with the colors of both nearby and distant galaxies 
\citep{munoz-mateos09,hainline09}. However, the  flux ratio $f_{24}/f_{70}=0.42$, while
inconsistent with an unobscured stellar source, is rather high for a galaxy detected at 70~\micron{};
such galaxies typically have  $0.05\lesssim f_{24}/f_{70} \lesssim 0.11$  \citep{frayer06}.
The flux ratios and colors are more consistent with those of candidate
young stellar objects detected in the Cepheus Flare \citep{kirk09}, although
YSOs do not typically have lenticular morphologies. Planetary nebulae can
have asymmetric shapes, but the SZ Tau emission has quite different colors (see Figure~\ref{fig:colors}).
The best candidates to explain the extended emission near SZ Tau are either an unusually-shaped  Galactic object
(possibly a YSO) or a background galaxy with very weak cold dust emission.
This mysterious object warrants further investigation. A more
definitive conclusion on its nature could be reached with higher spatial resolution infrared observations,
in which the flux from the star and extended emission were more clearly separated.

\subsection{T Mon and X Cyg}

\begin{figure}
\includegraphics[width=18cm]{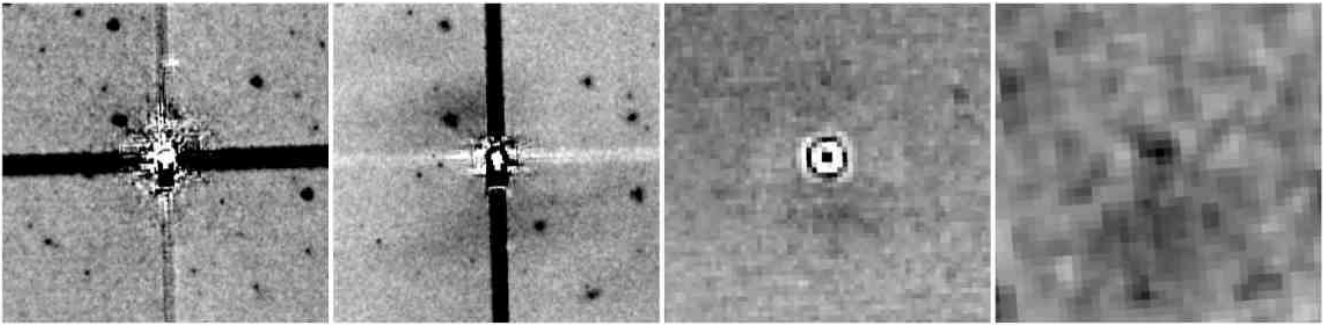}
\caption{PSF-subtracted images of T Mon, 
in the (left to right) 5.8 $\mu$m and
8.0 $\mu$m IRAC bands,
and the 24 $\mu$m  and 
70 $\mu$m  (not PSF-subtracted) MIPS bands,
All images are 2.5 arcminutes square.}
\label{fig-TMon}
\end{figure}

\begin{figure}
\includegraphics[width=18cm]{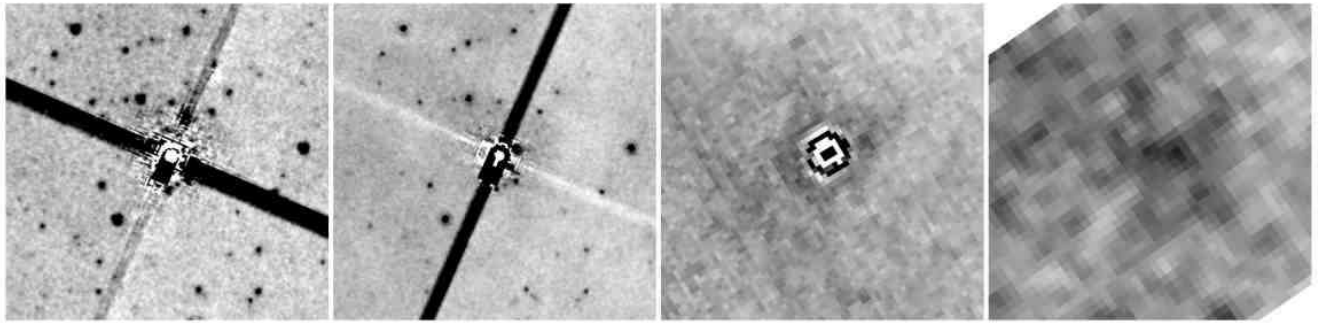}
\caption{
PSF-subtracted images of X Cyg in the 
(left to right) 5.8 $\mu$m and
8.0 $\mu$m IRAC bands,
and the 24 $\mu$m  and 
70 $\mu$m  (not PSF-subtracted) MIPS bands.
The field of view is a 2.5 arcminute square.
}
\label{fig-XCyg}
\end{figure}

These stars show extended emission in the IRAC 8.0~\micron{} band with only tentative
detections in the MIPS bands;
images can be seen in Figures \ref{fig-TMon} and  \ref{fig-XCyg}.  
In the image of  \object{T Mon}, there appears to be  cirrus across the entire field of view. 
However, there is brighter nebulosity closer to the star, specifically 
to the north and south in an hourglass-like shape extending to about 45\arcsec\ radius 
($6.2\times10^4$~AU, or 0.3~pc, at a distance of 1389~pc).
The emission is centered around the star, suggesting  that it may be locally illuminated.  
The extended emission near X~Cyg has the second-lowest significance (after GH Lup)
of the stars with emission above background levels. It is brightest to the northwest of the star
and concentrated within a radius of 30\arcsec\ ($3.5\times10^4$~AU, or 0.16~pc,
at a distance of 1163~kpc).  For both stars, the 24~\micron{} images show hints of enhanced
brightness at similar positions to the 8~\micron{} emission, but these positions are too
close to the PSF subtraction residuals to make definitive conclusions, or to measure
a surface brightness. There is no corresponding detectable emission at  70~\micron.
With both stars having extended emission observed at only 
one wavelength, the detections must be regarded as tentative. Observations with the
James Webb Space Telescope, particularly its coronographs, would provide an avenue
to confirm or refute them. If the detections are confirmed, 
the properties of these stars may provide a challenge in relating them to any mass loss.
\citet{neilson08}  predicted that mass loss rates, while showing large scatter with period, are generally larger for
long-period stars. T Mon is a long period star with a companion \citep{evans99}, while X Cyg 
has a period of 16.4~d and no known companion hotter than type A5 \citep{evans92}.
However, a low-mass companion with an orbit in the plane of the sky is not ruled out
for X~Cyg and might be important in forming and/or shaping an extended nebula.

\section{Discussion and Conclusions}
\label{sec:disc}

{\em Spitzer\/}{} mid-infrared observations of Cepheids have produced no strong
evidence for large amounts of warm circumstellar dust \citep{marengo10a}. 
Here we use the same observations to place limits on cooler dust. Of our sample
of 29 Cepheids, two stars (RS~Pup and $\delta$~Cep) have well-detected extended infrared emission,
two (GH Lup and S Mus) have possible extended emission at 24 and 70~\micron{}, and a further three
($\ell$~Car,  T Mon and X Cyg) show tentative evidence for extended infrared
emission at shorter infrared wavelengths only. 

Is there anything particularly special about the stars in which we do have
(tentative) evidence for mass loss? 
Our observations of the nebulosity around $\delta$~Cep and S~Mus suggest that 
the detectability of a dustless wind might be enhanced when interstellar matter around the 
Cepheid is swept up into a bow shock, and a hot companion heats the material.
Compared to the emission around RS~Pup and GH Lup, the $S_{24}/S_{70}$ ratio (a proxy
for dust temperature) is higher near S~Mus and $\delta$~Cep.
Both of these stars have known hot companions, while RS Pup and GH Lup do not
(although low-temperature companions cannot be excluded using current data).
A counter example might be provided by T Mon, which also has a hot companion, but 
does not have detectable extended emission at 24 or 70~\micron{} or an infrared excess;
however, T Mon's companion is cooler (type A0) than that
of the other two stars (B3 and B7-8).

The  surface brightness ratios $S_{8.0}/S_{24}$ are a rather crude
tool for deriving detailed physical conditions, but can give
some insight into the PAH content of the ISM near these stars.
The low PAH content indicated by the $S_{8.0}/S_{24}$ ratios for RS~Pup, S~Mus and $\delta$~Cep---much lower than
in the NGC 7023 reflection nebula, or the ISM in general---suggests that
the ISM around these stars has been diluted to some extent by the stellar winds. 
The higher PAH content near GH Lup would imply less dilution, suggesting that the
emission there may be unrelated to the star. We caution that, while
the amount of dilution might be indicated by the
$S_{8.0}/S_{24}$ ratio, high spatial-resolution spectroscopy in which the stellar and
extended emission are separated would provide much greater physical
insight into the conditions near these stars.
 
The IRAC-only extended emission near $\ell$~Car, T Mon, and X Cyg also
suggests that these stars are somehow unusual.  However, we have been unable to
find any clear distinction between these stars and the three which also show
extended MIPS emission, or between all seven stars and our full Cepheid sample.
Like RS Pup,  $\ell$~Car and T Mon have long periods ($>27$~d) and low temperatures; 
GH~Lup,  S~Mus and $\delta$~Cep have periods $<10$~d, and X~Cyg is intermediate
with a period of 16.4~d.
Six of the seven stars have super-solar metallicity \citep{romaniello08, andrievsky05,kovtyukh05}
but the difference in mean metallicity between these stars and the other stars
in the sample is barely statistically significant ($p=0.025$).
\citet{evans92} placed strong limits on the presence of hot companions in Cepheids:
 of the stars in our sample, 3 of the 7 with possible extended emission have such companions
(S Mus, $\delta$ Cep, T Mon), as do 5 of the 22 stars without extended emission 
(an additional 3/23 are radial-velocity binaries). With such small numbers it is
not yet possible to say whether the presence of hot companions enhances
the detection of extended infrared emission.

RS Pup is clearly different from all of the other Cepheids in our sample in terms of the extent
and brightness of its infrared emission. One possible explanation is that this star has swept
up more material, either by having been on the instability strip for longer (unlikely, since
long-period stars evolve faster) or by being
located in a region of higher density in the interstellar medium.
An alternative explanation is suggested by visual examination of a Schmidt plate taken by one of us
(DW) in the region of RS Pup. The image reveals a number of near-identical dust clumps, 
with similar angular sizes, in the vicinity of the RS Pup nebula. If these clumps are
at the same physical distance and are physically related, but only one of them has a Cepheid in it, 
it could be argued that  they are all likely pre-existing nebulae---that is, not the result of Cepheid mass loss.
The counter-argument  would be that the ring-like structures observed in the
surrounding nebula appear to be centered on  RS Pup \citep{havlen72}: this would
be an extreme coincidence if the nebula were unrelated to the star.
 
Before our {\em Spitzer}{} observations, only two Cepheids had unambiguous evidence for 
circumstellar nebulae: $\ell$ Car and RS Pup. We have now doubled that number
with the addition of S~Mus and $\delta$ Cep; if extended emission around
GH Lup, T~Mon and X~Cyg is confirmed, this would increase the sample size even more.
A rough estimate of the  fraction
of Cepheids with substantial dusty mass loss can be made as 4 to 7 of 29, or  $ \sim 14-24$\%.%
\footnote{The shells detected in the near-infrared around some Cepheids \citep{merand06,merand07,kervella06}
are not consistent with dusty mass loss: they are very close to the stars 
and thus well above the dust sublimation temperature \citep{marengo10a}.}
The true fraction of Cepheids with mass loss could be below this limit if the observed nebulosity is related
 to the gas and dust from which the star formed. This is however unlikely, because even for small 
 space velocities ($\sim 1$~km~s$^{-1}$, an order of magnitude less than the observed velocity of $\delta$ Cep and $\ell$ Car), 
 in 50 Myr (the time it takes for an intermediate mass star to become a Cepheid) these stars will have moved 
 away from their birthplace by several tens of parsecs. This is much larger than the size of the observed nebulosity around the stars in our sample.

Addressing the initial motivation for this study, it appears that mass loss, at least its dusty component visible in the infrared, 
can only partially account for the Cepheid mass loss discrepancy 
\citep[e.g.\ up to 1/5th of a 10\% discrepancy in the case of $\delta$~Cep, see][]{marengo10b}. 
Our estimates of Cepheid mass loss (in the range $10^{-8}$ to $10^{-9}$~M$_{\sun}$~yr$^{-1}$) 
are comparable to previous studies based on the detection of infrared and UV excess \citep[e.g.][]{welch88, deasy88}, 
but the higher spatial resolution of our images helps to separate local emission from background contamination. 
The detection of shocked structures, as in the case of $\delta$~Cep and possibly $\ell$~Car, is a compelling indication 
of current mass loss in those stars. While the total mass lost over the Cepheid phase is small, the confirmation 
that these processes are active needs to be taken into account for a proper modeling of Cepheid evolution. 
The presence of significant Cepheid winds can affect hydrodynamic models of Cepheid pulsations, and the 
way shock waves propagate in the Cepheid atmospheres. Finally, the observed level of circumstellar dust 
needs to be further explored, as it may have a significant effect on inferred Cepheid distances as shown by \citet{neilson10}.

\acknowledgments

The authors thank H. Neilson for useful conversations and L.D. Matthews
for assistance with computing the projected space velocity of $\ell$ Car.
We thank the referee for helpful comments.
This work is based on observations made with the \textit{Spitzer Space Telescope},
which is operated by the Jet Propulsion Laboratory, California Institute of 
Technology under NASA contract 1407. 
Support for this work was provided by NASA through an award issued by JPL/Caltech.
PB and DLW acknowledge research support through Discovery Grants from the Natural Sciences
and Engineering Research Council of Canada; PB also acknowledges support
from an Ontario Early  Researcher Award.
NRE acknowledges support from the Chandra X-Ray Center grant NAS8-03060.
DH was supported in part by the National Science Foundation REU
and Department of Defense ASSURE
programs under Grant no. 0754568 and by the Smithsonian Institution.

{\it Facilities:} \facility{Spitzer (IRAC, MIPS)}.

\bibliographystyle{aas}

\begin{deluxetable}{lllll}
\tabletypesize{\small}
\tablewidth{0pt}
\tablecaption{Surface photometry of extended infrared emission
\label{tab:surfphot}}
\tablehead{\colhead{Star/region}  & \colhead{5.8~\micron{}} & \colhead{8~\micron{}} & \colhead{24~\micron{}} & \colhead{70~\micron}}
\startdata
GH Lup 1 & $0.07\pm0.01$ &$0.27\pm0.01$  &$0.16\pm0.02$ &$1.1\pm0.1$ \\
GH Lup 2 &  $0.02\pm0.01$& $0.36\pm0.01$ & $0.29\pm0.02$& $1.8\pm0.1$\\
GH Lup 3 &  $0.13\pm0.01$&  $0.49\pm0.01$& $0.51\pm0.02$& $3.5\pm0.2$\\
GH Lup 4 &  $0.10\pm0.01$& $0.41\pm0.01$ & $0.46\pm0.02$& $2.6\pm0.1$\\
RS~Pup 1 & $0.24\pm0.03$ &$0.63\pm0.02$  & $0.89\pm0.39$& $9.4\pm0.8$\\
RS~Pup 2 &  $0.18\pm0.01$& $0.56\pm0.01$ & $0.82\pm0.30$& $24\pm1$\\
RS~Pup 3 &  $0.27\pm0.01$& $0.83\pm0.03$& $1.99\pm0.86$& $40\pm1$\\
RS~Pup 4 &  $0.33\pm0.01$&  $0.91\pm0.01$& $3.10\pm0.72$& $47\pm1$\\
RS~Pup 5 &  $0.23\pm0.01$&  $0.84\pm0.03$& $3.14\pm2.08$& $35\pm2$\\
S~Mus 1\tablenotemark{a} & $0.05\pm0.01$ &$0.10\pm0.01$  &$1.3\pm1.1$ &$2.0\pm0.2$ \\
S~Mus 2 &  $0.08\pm0.01$& $0.09\pm0.01$ & $0.8\pm0.6$& $1.9\pm0.1$\\
S~Mus 3 &  $0.15\pm0.01$&  $0.16\pm0.01$& $0.7\pm0.3$& $0.9\pm0.1$\\
S~Mus 4 &  $0.07\pm0.01$& $0.11\pm0.01$ & & \\
\enddata
\tablenotetext{a}{For S Mus, regions are not the same in IRAC and MIPS.}
\tablecomments{All surface brightness values in MJy~sr$^{-1}$.}
\end{deluxetable}

\clearpage

\end{document}